\documentclass[aps,prl,reprint,groupedaddress,amsmath,amssymb]{revtex4-1}
\usepackage{graphicx}
\usepackage{epstopdf}
\usepackage{subfigure}
\usepackage{dcolumn}
\usepackage{bm}
\usepackage{hyperref}
\hypersetup{colorlinks=true,       
    linkcolor=blue,          
    citecolor=blue,        
    urlcolor=blue }	

\begin{document}
\title{Beating the standard quantum limit for  force sensing \\ with a coupled two-mode optomechanical system}


\author{Xunnong Xu }
\author{Jacob M. Taylor}

\affiliation{Joint Quantum Institute, University of Maryland/National Institute of Standards and Technology, College Park, Maryland 20742, USA}


\date{\today}

\begin{abstract}
Optomechanics allows the transduction of weak forces to optical fields, with many efforts approaching the standard quantum limit.  We consider force-sensing using a mirror-in-the-middle setup and use two coupled cavity modes originated from normal mode splitting for separating pump and probe fields. We find that this two-mode model can be reduced to an effective single-mode model, if we drive the pump mode strongly and detect the signal from the weak probe mode. The optimal force detection sensitivity at zero frequency (DC) is calculated and we show that one can beat the standard quantum limit by driving the cavity close to instability. The best sensitivity achievable is limited by mechanical thermal noise and by optical losses. We also find that the bandwidth where optimal sensitivity is maintained is proportional to the cavity damping in the resolved sideband regime. Finally, the squeezing spectrum of the output signal is calculated, and it shows almost perfect squeezing at DC is possible by using a high quality factor and low thermal phonon-number mechanical oscillator. 
\end{abstract}


\maketitle

Dramatic progress in coupling mechanics to light \cite{Kippenberg2008, Marquardt2009, Milburn2011, Aspelmeyer2012} suggests that such devices may be used in a wide variety of settings to explore quantum effects in macroscopic systems.  Furthermore, such systems can be exquisitely sensitive to small perturbations, such as forces induced either by acceleration as in accelerometer \cite{Krishnan2007} or by, e.g., coupling to surfaces or fields as in atomic force microscopy \cite{Binning1986}.  For such force measurements, a high quality factor (Q) mechanical oscillator acts as a test mass, transducing a force into a time-dependent displacement of the oscillator \cite{Braginsky1992, Caves1980}.  By using interferometric techniques to monitor the position of the oscillator, one can infer the force via optical signals.  However, the radiation pressure coupling between the mechanical mode and optical mode has three consequences: photon shot noise, quantum backaction and dynamical backaction \cite{Kippenberg2008, Clerk2010}. The dynamical backaction modifies the oscillator dynamics \cite{Sheard2004} and makes laser cooling \cite{Arcizet2006nature, Schliesser2008} or amplification of phonons \cite{Bagheri2011} in the mechanical system possible. Photon shot noise and quantum backaction, the former decreases with increasing input laser power while the latter increases with increasing input laser power, introduce two sources of noise on the displacement readout of the oscillator motion. An optimal compromise between these two noise sources leads to the standard quantum limit (SQL) in force sensing \cite{Braginsky1992}.  

The SQL, however, is itself not a fundamental limit. By using squeezed states of light \cite{Xiao1987},  employing quantum nondemolition (QND) measurement \cite{Thorne1978}, or by cavity detuning \cite{Arcizet2006}, the SQL can be surpassed.  Here we show that in a coupled two-mode optomechanical system, if we drive it appropriately, the interaction between cavity photons and the mechanical oscillator will generate squeezed states of the output light. Measuring an appropriate quadrature of the output light field, we would get fewer fluctuations than that of the vacuum state, which makes it possible to detect weak forces far below the SQL. Furthermore, since we pump and probe different resonant optical modes, the effective optomechanical coupling is enhanced, and thus the pump power requirement for achieving the best sensitivity is lowered substantially.

\begin{figure}[!t]
\includegraphics[width = \columnwidth]{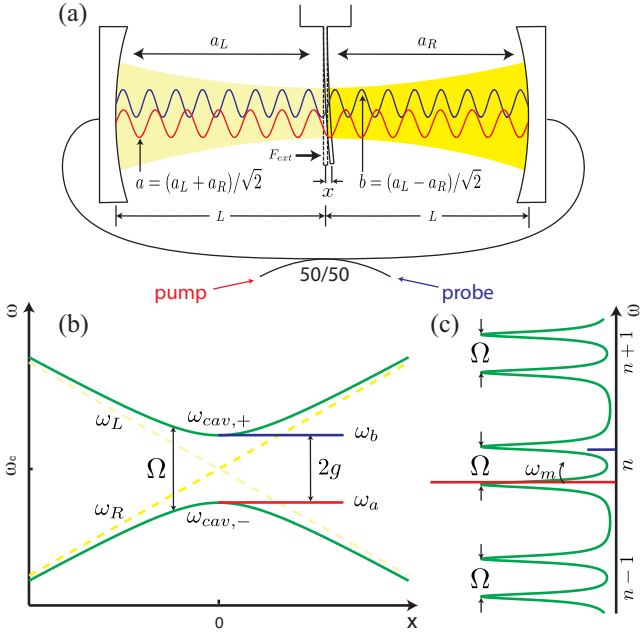}
\caption{\label{fig1}(Color online). (a) A symmetric mirror-in-the-middle optomechanical system comprising a Fabry-Perot cavity of length $2L$ with a high reflectivity mirror mounted in the middle and coupled to a mechanical oscillator.  Displacements of the middle mirror (via mechanical oscillations) couple the two normal modes $a$ (red) and $b$ (blue) as the left-right symmetry is broken. (b) Normal mode frequencies $\omega_{cav,\pm}$ (green) as a function of middle mirror displacement $x$. (c) Transmission spectrum of the three-mirror cavity, showing pairs of normal modes. We drive mode $a$ strongly (long red line) and detect mode $b$ (short blue line). }
\end{figure}

We consider a high finesse Fabry-Perot cavity with a dielectric mirror in the middle \cite{Thompson2008,Jayich2008} [Fig.~\hyperref[fig1]{1(a)}].
The end mirrors of the FP cavity are fixed, while the middle mirror can vibrate along the optical axis of the cavity at a mechanical frequency $\omega_m$ with effective mass $m$. With the presence of the middle mirror, the FP cavity is divided into two sub-cavities, denoted by left (L) cavity  and right (R) cavities. The middle mirror has a nonzero transmission, which allows the exchange of light between these two subcavities and thus leads to an effective coupling between the left and right cavity modes \cite{Marquardt2011}. Further, the coupling will shift the resonant frequencies of  two coupled cavity modes and leads to the so-called normal mode splitting effect \cite{Agarwal1984, Thompson1992}.

Following \cite{Bhattacharya2008, Jayich2008}, the normal mode splitting in the presence of the middle mirror can be calculated by assuming a transfer matrix with a high reflectivity $r_d$. For simplicity, we assuming the middle membrane to be exactly at the middle point of the FP cavity initially, dividing the cavity into two subcavities with the same length $L$, and the normal mode splitting when the middle mirror is at a new position $x$ is given by $\Omega = \cfrac{c}{L} \arccos(|r_d|\cos(2kx))$ [Fig.~\hyperref[fig1]{1(b)}], where $r_d$ is complex amplitude reflectivity of the middle mirror, $k$ is the wave vector of the incoming field and $x$ is the displacement from the middle point of the FP cavity. 

The hamiltonian of the cavity fields in this three-mirror system is
\allowdisplaybreaks
\begin{eqnarray}
H_{\mathrm{opt}}  & = &   \hbar (\omega_c +   f  x )  a_L^{\dagger} a_L  + \hbar  (\omega_c  -  f   x ) a_R^{\dagger} a_R   \nonumber \\
&&- \hbar g(a_L^{\dagger}a_R + a_R^{\dagger}a_L),
\label{eq2}
\end{eqnarray}
where $\omega_c$ is the resonance frequency of the  subcavities with the middle mirror exactly in the middle ($x=0$),  $f = \left.(\partial \omega_{cav}/\partial x) \right |_{x=0}$ is the shift of cavity resonant frequency per unit length evaluated at $x=0$, and $a_L(a_R)$ is the left (right) cavity mode annihilation operator.  Here only the linear order frequency shift is considered, since the displacement $x$ is much smaller than the cavity length $L$. The last term describes the coupling between left/right cavity modes with strength $g$. In the high reflectivity limit ($|r_d| \rightarrow 1$), $g = |t_d| c/2L$ and $ f = - \omega_c/L$, where $t_d$ is the amplitude transmission coefficient and $c$ is the speed of light in vacuum. 


We look only in a narrow spectral range around a nominal pair of normal modes  $a= (a_L + a_R)/\sqrt{2}$ and $b= (a_L - a_R)/\sqrt{2}$. We also drive mode $a$ strongly at frequency $\omega_L$ and move to the rotating frame with respect to the pump laser \cite{Scully1997}. The cavity field is coupled to fields outside the cavity through the ends mirrors, while we assume the mechanical oscillator is coupled to a thermal bath through clamping losses. The classical and quantum fluctuations of the environmental degrees of freedom will introduce damping to the cavity field and mechanical oscillator \cite{Gardiner2004, Walls2008}, as required by the fluctuation-dissipation theorem \cite{Kubo1966}. In the Markovian approximation, the Heisenberg-Langevin equations for mechanical and optical degrees of freedom are, in the high temperature limit, as follows: 
\begin{subequations}
\begin{eqnarray}
\dot{x} & = &p/m, \\
\dot{p} & = & -m\omega_m^2 x -\gamma p + \frac{\hbar \omega_c}{L} (a^{\dagger} b + b^{\dagger} a) + F_{in}, \\
\dot{a} & = & - i(\Delta_c-g)  a - \kappa a + i \frac{\omega_c}{L} b x + E + \sqrt{2\kappa}a_{in},\\
\dot{b} & = & - i(\Delta_c+g)  b - \kappa b + i \frac{\omega_c}{L} a x  + \sqrt{2\kappa}b_{in}. 
\end{eqnarray}
\end{subequations}
In the equations above, $\gamma$  is the damping of the mechanical oscillator, $\kappa$ is the damping of the cavity, and $\Delta_c = \omega_c - \omega_L$ is the cavity detuning. $E$ is the pump strength which is related to input laser power $P_{in}$  and cavity damping $\kappa$  by $|E| = \sqrt{P_{in}\kappa/\hbar\omega_L}$. $a_{in}$, $b_{in}$ are the vacuum fluctuations of the two cavity modes. $F_{in}$ is the force acting on the oscillator, and it has two parts: an external force $F_{ext}$ acting on the oscillator, which is also the force to be detected [Fig.~\hyperref[fig1]{1(a)}]; the Brownian stochastic force, or thermal fluctuating force $F_{th}$, which leads to damping of the oscillator.

We then find the steady state of the system to be  $\langle x \rangle= 0$,  $\langle p \rangle= 0$, $\langle b\rangle = \beta =  0$, $\langle a\rangle = \alpha = E/[i (\Delta_c -g) + \kappa]$, which is the only stable solution at low power. Following \cite{DeJesus1987, Genes2008}, we study the stability of this solution by applying the Routh-Hurwitz criterion \cite{Routh-Hurwitz}, which at positive effective detuning $\Delta = \Delta_c + g >0$ is simplified to a constraint on pump strength, 
\begin{equation}
\alpha^2 < m\omega_m^2 L^2 (\kappa^2+\Delta^2)/2\hbar\Delta\omega_c^2 = \alpha_0^2.
\label{stability_condition}
\end{equation}
Writing $x, p, a, b$ in terms of steady state value and fluctuation, and neglecting terms of order $\cfrac{\omega_c}{L} \tilde{b}\tilde{x}$, $\cfrac{\omega_c}{L} \tilde{a}\tilde{x}$, $\cfrac{\hbar \omega_c}{L} \tilde{a}\tilde{b}$, then the fluctuation of mode $a$ decouple, and the equations of motion reduce to,  
\begin{subequations}
\begin{eqnarray}
\dot{x} & = &  p/m, \\
\dot{p} & = & -m\omega_m^2 x -\gamma p + \hbar G (b + b^{\dagger}) + F_{in}, \\
\dot{b} & = & - i\Delta  b - \kappa b + i G x  + \sqrt{2\kappa}b_{in},
\end{eqnarray}
\end{subequations}
where we choose $\alpha$ to be real and define $G = \omega_c\alpha/L$. For convenience, we remove the tilde of the fluctuating variables. We find that the pumped, coupled two-mode model is reduced to an effective single-mode model \cite{Borkje2012}, where the cavity  consist of a fixed partial transmitting mirror and movable perfect reflecting mirror and the optomechanical coupling strength is determined by $\alpha$, the strength of pump.

We then define the quadratures of mode $b$ as $X =  (b + b^{\dagger})/\sqrt{2}$,  $Y = (b - b^{\dagger})/i \sqrt{2}$, move to frequency domain by Fourier transform,  and solve a set of linear equations. We find 
\begin{eqnarray}
x(\omega) &=& \chi(\omega)   F_{in}(\omega)  + \chi(\omega) \cfrac{2\hbar G\sqrt{\kappa}}{(\kappa-i\omega)^2 +\Delta^2} \nonumber \\
&& \times \left [(\kappa-i\omega) X_{in}(\omega) + \Delta Y_{in}(\omega) \right],
\label{eq12}
\end{eqnarray}
where $\chi(\omega)$ is the  susceptibility of the optomechanical system to force, 
\begin{equation}
\chi(\omega) = \left\{m\left [\omega_m^2 -  \omega^2 -i\gamma \omega  -\cfrac{2\hbar G^2\Delta/m}{(\kappa-i\omega)^2 +\Delta^2} \right] \right\} ^{-1} . 
\end{equation}
From the expression above, we immediately identify an effective, frequency-dependent mechanical resonant frequency $\omega_m^{\prime}$ and an effective damping $\gamma^{\prime}$ which are shifted from the original ones. The shift in resonant frequency is the so called ``optical spring" effect \cite{Sheard2004}, while the shift in damping  leads to cooling or heating of the oscillator, depending on the sign of detuning \cite{Kippenberg2008, Marquardt2009}. The cavity field fluctuations enter the equation of motion for oscillator Eq.~(\ref{eq12}) as an additional fluctuations force, which is identified as the shot noise fluctuations of radiation pressure force.

Within the input-output formalism \cite{Walls2008}, the output field quadratures we measure are related to the field quadratures inside the cavity by $X_{out}(\omega) = \sqrt{2\kappa} X(\omega) - X_{in}(\omega)$ and $Y_{out}(\omega) = \sqrt{2\kappa} Y(\omega) - Y_{in}(\omega)$. 
We consider a homodyne measurement of the signal  \cite{Fermani2004, Walls2008}
\begin{eqnarray}
S(\omega) & = & \sin\theta X_{out}(\omega) + \cos\theta Y_{out}(\omega) \\
& = & \chi_F(\omega) F_{in}(\omega) + \chi_X(\omega)  X_{in}(\omega) + \chi_Y(\omega) Y _{in}(\omega), \nonumber 
\end{eqnarray}
where $\theta$ is an experimentally adjustable phase, which determines the measured quadrature.  Here the signal is written in terms of three inputs (force input $F_{in}$, amplitude fluctuations $X_{in}$, phase fluctuations $Y_{in}$) and corresponding susceptibilities. The force and field susceptibilities are: 
\begin{subequations}
\begin{eqnarray}
\chi_F(\omega)& = &  \cfrac{2\sqrt{\kappa} G [\Delta \sin\theta + (\kappa-i\omega)\cos\theta ]}{(\kappa-i\omega)^2 + \Delta^2}\chi(\omega)  \\
\chi_X(\omega) & = &  \cfrac{4 \hbar \kappa G^2 [\Delta \sin\theta + (\kappa-i\omega)\cos\theta ] (\kappa-i\omega)}{[(\kappa-i\omega)^2 +\Delta^2]^2} \chi(\omega) \nonumber \\
&&  + \frac{(\kappa^2 + \omega^2 -\Delta^2)\sin\theta - 2\kappa\Delta\cos\theta}{(\kappa-i\omega)^2 + \Delta^2} , \\
\chi_Y(\omega) & = & \cfrac{4 \hbar \kappa G^2 [\Delta \sin\theta + (\kappa-i\omega)\cos\theta ] \Delta}{[(\kappa-i\omega)^2 +\Delta^2]^2} \chi(\omega) \nonumber \\
&& + \frac{2\kappa\Delta\sin\theta + (\kappa^2 + \omega^2 -\Delta^2)\cos\theta}{(\kappa-i\omega)^2 + \Delta^2}.
\end{eqnarray}
\end{subequations}

To calculate the sensitivity to the external force $F_{ext}$, we define the following quantity 
\begin{eqnarray}
F(\omega) & = &  \left.  \frac{S(\omega)}{\partial S(\omega)/ \partial F_{ext}} \right  |_{F_{ext} = 0 } ,
\end{eqnarray}
and the square of force detection sensitivity is given by the power spectral density of $F(\omega)$: $\eta(\omega)= \int d\omega^{\prime} \langle F(\omega) F(\omega^{\prime} ) \rangle.$
The vacuum radiation input noise $b_{in}$ is delta correlated and the thermal fluctuating force is approximated as white noise thus is also delta correlated, so we have
\begin{equation} 
\eta(\omega) =  2m\gamma k_B T + \frac{1}{2}\left| \frac{\chi_X(\omega)  - i \chi_Y(\omega)}{\chi_F(\omega)}\right|^2 .
\label{sensitivity}
\end{equation}
The first term is the thermal white noise. There is a term proportional to $G^2$ which comes from the random back action force, and also a term proportional to $1/G^2$ which is the phase noise related to position measurement imprecision.

We now focus on the DC ($\omega =0 $) force sensing regime. To get the best sensitivity, we first optimize the function $\eta(\omega = 0)$ for $\alpha$ and then optimize for $\theta$. We find that for the optimal pump strength $\alpha_{\ast}^2$, the second term goes to zero as  $\Delta\sin\theta + \kappa\cos\theta \to 0$, which corresponds to a backaction free point. At first sight, it seems that $\chi_F$ approaches zero as $\Delta\sin\theta + \kappa\cos\theta \to 0$, and the sensitivity diverges. However, if we choose the pump strength appropriately, then not only can the divergence at $\theta_0 = -\arctan(\kappa/\Delta)$ be avoided, but also the sensitivity can achieve its optimal value. The optimal pump strength is given by  
\begin{equation}
\alpha_{\ast}^2 =  \alpha_0^2 \left(1 - \frac{2\kappa}{\Delta}\frac{\Delta\sin\theta + \kappa\cos\theta}{\sqrt{\kappa^2 + \Delta^2}}\right),
\label{optimal_pump}
\end{equation}
where $\alpha_0$ is the threshold pump strength defined in Eq.~(\ref{stability_condition}). At this point, the effective mechanical frequency $\omega_m^{\prime} \to 0$ as $\theta \to \theta_0$, so $\chi_F$ is still finite. We then try to find out the behavior of the sensitivity near the critical angle $\theta_0$. To ensure that the pump strength does not exceed the threshold value and that the effective mechanical frequency is positive,  we let the angle $\theta$ approaches $\theta_0$ from the positive side, that is $\theta = \theta_0 + \delta\theta$, with $0< \delta\theta \ll 1$. We note that, if $\theta$ approaches $\theta_0$ from the negative side, we can replace the minus sign in Eq.~(\ref{optimal_pump}) with a plus sign, and the result will be similar. In the vicinity of $\theta_0$, the pump is approximated as 
\begin{equation}
\alpha_{\ast}^2  = \alpha_0^2 (1 - \frac{2\kappa}{\Delta}\delta\theta).
\end{equation}

At the optimal pump strength and optimal angle, the total sensitivity is found:
\begin{equation}
\eta(\omega = 0) 
\approx   2m\gamma k_B T + \hbar m\omega_m^2 \left(\frac{\Delta}{4\kappa} + \frac{\kappa}{\Delta}\right)\xi^2,
\label{DC_sensitivity}
\end{equation}
with the dimensionless parameter $\xi$ defined by $\xi = \delta\theta\cdot2\kappa/\Delta$. This result implies that thermal noise limited detection can be achieved by choosing the critical angle $\theta_0$ appropriately. The expense for achieving the best sensitivity is that we have to pump the system at a power close to the threshold value, thus increasing the possibility of destabilizing the system. The ultimate sensitivity for force detection is limited by how strong the thermal noise is and how close we can pump the system near its instability point. The result also suggests that we could further improve the sensitivity by choosing $\Delta = 2\kappa$. Then we look at the input laser power, and find that if $\Delta_c = g$, the power is minimized. Along with the condition $\Delta = 2\kappa$,  we have $\Delta_c = g=\kappa$, and the optimal pump power is 
\begin{equation}
P_{\mathrm{opt}} 
 \approx  \frac{5}{4}(1 - \xi) m\omega_m^2\left(\frac{L}{Q_c}\right)^2\omega_c,
\end{equation}
where $Q_c = \omega_c/\kappa $ is the quality factor of the cavity. Considering an optomechanical system with $m = 5.36\times 10^{-10}~\mathrm{Kg}$, $\omega_m = 2\pi\times 130~\mathrm{kHz}$, pumping laser wavelength $\lambda = 1.55~\mathrm{\mu m}$, and cavity finesse of $F = 20000$ \cite{Kemiktarak2012}, we find that for $\xi =0$, $P_{\mathrm{opt}} = 0.816~\mathrm{mW}$, and the circulating power $P_{\mathrm{cir}} = P_{\mathrm{opt}}F/2\pi = 2.56~\mathrm{W}$.

In practice,  we need to understand the behavior of $\eta(\omega)$ at low but nonzero frequencies to determine the bandwidth for force detection. A full analysis is only possible numerically, so here we present a simple but illuminating approximation method. The argument is the following: at nonzero frequencies, in order to keep the $\delta\theta$ dependence in Eq.~(\ref{DC_sensitivity}), we require $|\omega\cos\theta| < \Delta\sin\theta + \kappa\cos\theta$, which is equivalent to 
\begin{equation}
|\omega| < \frac{1}{2}\left[1 + \left(\frac{\Delta}{\kappa}\right)^2\right] \xi\kappa.
\label{bandwidth}
\end{equation}
Thus the bandwidth is approximately $\left[1 + (\Delta/\kappa)^2\right]\xi\kappa$. 
\begin{figure}[h!]
\includegraphics[width = \columnwidth]{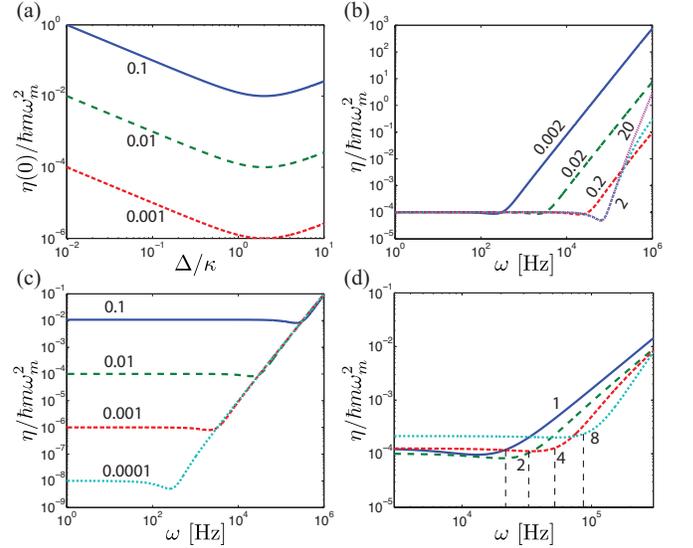}
\caption{\label{fig2}(Color online). (a) Sensitivity at DC as a function of $\Delta/\kappa$ for different values of $\xi$ (numerics shown in plot). At $\Delta/2\kappa$, the minimum value is achieved. In the following figures,  we choose $\omega_m = 2\pi\times 10^6~\mathrm{Hz}$, $\kappa = 0.2\omega_m$, $\Delta = 2\kappa$ and $\xi = 0.01$ as the base values to plot the sensitivity as a function of frequency. The bandwidth is the frequency range where the best sensitivity is maintained. (b) Bandwidth dependence on $\kappa$, where $\omega_m$ is fixed and $\kappa$ varies and the ratio $\kappa/\omega_m$ is shown in plot. In the resolved sideband regime the bandwidth increases with $\kappa$ linearly, while in the unresolved regime, it does not increase much.  (c) Bandwidth dependence on $\xi$, which is approximately linear. (d) Bandwidth dependence on $\Delta/\kappa$ , where $\kappa$ is fixed and $\Delta$ varies. The dependence is approximately in the form $\sqrt{1 + (\Delta/\kappa)^2}$ and suggests a narrower bandwidth than our estimate Eq.~(\ref{bandwidth}).}
\end{figure}
Neglecting the thermal noise term, we plot the DC sensitivity for different values of $\xi$ in Fig.~\hyperref[fig2]{2(a)}, and the sensitivity at low signal frequencies for different values of $\kappa$, $\xi$, $\Delta/\kappa$ in Fig.~\hyperref[fig2]{2(b)}, \hyperref[fig2]{2(c)}, \hyperref[fig2]{2(d)} respectively. We find that the bandwidth shown in the numerical result is in good agreement with Eq.~(\ref{bandwidth}). Thus we could use this formula to estimate the bandwidth at low frequencies for a given set of parameters.

Finally, we study the squeezing spectrum of the output signal $S(\omega)$, which is given by $\tilde{S}(\omega) = \int \mathrm{d}\omega^{\prime} \langle S(\omega), S(\omega^{\prime}) \rangle$ \cite{Collett1984, Walls2008}. Using the noise correlation relations, we have  
\begin{equation}
\tilde{S}(\omega) = 2m\gamma k_B T \left| \chi_F(\omega) \right|^2 + \frac{1}{2} \left| \chi_X(\omega) - i\chi_Y(\omega) \right|^2.
\end{equation}
At DC, we can minimize it by choosing the optimal angle and pump strength, obtaining
\begin{equation}
\tilde{S} 
= \frac{n_{th}}{Q_m}\frac{\Delta}{\kappa}( 1-\xi) +  \frac{1}{2}\left[ 1 + \left(\frac{\Delta}{2\kappa}\right)^2\right] \xi^2. 
\end{equation}
where $n_{th} = k_B T/\hbar\omega_m$ is the number of thermal phonon and $Q_m = \omega_m/\gamma$ is the quality factor of the mechanical oscillator. For $\Delta = 2\kappa$, $\tilde{S} = 2( 1-\xi)n_{th}/Q_m +  \xi^2$. Nano-mechanical oscillators of high quality factor and low phonon number have now been fabricated by many groups, which makes it possible to reduce the thermal noise term to a very small value. By driving the system near the threshold ($\xi \to 0$), squeezing ($\tilde{S} < 1$) in the signal we measure could be realized. This is the optomechanical analog of the squeezing in the output field from an optical parametric oscillator (OPO) \cite{Scully1997}. Similarly, choosing $\theta \to \pi/2 + \theta$, we calculate the squeezing spectrum of a signal $\pi/2$ out of phase and find that the optical noise term is proportional to $1/\xi^2$, which is consistent with the Heisenberg uncertainty relation. 
 
In all the analysis above, however, we have not yet taken into account the quantum noise $X^{\prime}_{in}(\omega)$ introduced when measuring the field quadrature $S(\omega)$. Considering a measurement efficiency $P <1$, then the actual signal we measure is $S^{\prime}(\omega) = \sqrt{P}S(\omega) + \sqrt{1-P}X^{\prime}_{in}(\omega)$. We find the modified sensitivity to be $\eta^{\prime}(\omega) = \eta(\omega)  + (1-P)/2P|\chi_F(\omega)|^2$. Assuming that the optimization at  $\alpha_{\ast}^2$ and $\theta_0$ is still valid, then the sensitivity at DC is  $\eta^{\prime}(0) = \eta (0)  + \hbar m\omega_m^2 (1-P)/(1-\xi)\cdot \kappa/\Delta$, and the corresponding squeezing spectrum is $\tilde{S}^{\prime} = \tilde{S} + (1-P)/2$, in the limit $P \to 1$. Thus, a homodyne measurement efficiency of $99\%$ will limit the squeezing to $23~\mathrm{dB}$ (decibel). 

Our approach to squeezing for improved force sensing may also have direct application in related topics, including atomic force microscopy, magnetic resonance force microscopy, and even in quantum transduction via mechanics as recently suggested~\cite{PeterRabl2011}.  Furthermore, more complicated cavity mode structures, such as those of higher orbital angular momentum, may provide additional methods for achieving this outcome in single-side cavities, as the fundamentals of our approach are simply having two well isolated, near-by cavity modes that both interact with the mechanical degree of freedom.

We thank J. Harris, P. Meystre, J. Kimble, and C. Caves for helpful discussions.  JMT acknowledges KITP at UCSB for their hospitality during the quantum control meeting.  Funding is provided by DARPA QuASAR and the NSF Physics Frontier Center at the JQI. This research was supported in part by the National Science Foundation under Grant No. NSF PHY11-25915 and Grant No. PHY05-25915.

\nocite{*}
\bibliography{draft_v10_arXiv}

\providecommand{\noopsort}[1]{}\providecommand{\singleletter}[1]{#1}%
\begin{thebibliography}{35}%
\makeatletter
\providecommand \@ifxundefined [1]{%
 \@ifx{#1\undefined}
}%
\providecommand \@ifnum [1]{%
 \ifnum #1\expandafter \@firstoftwo
 \else \expandafter \@secondoftwo
 \fi
}%
\providecommand \@ifx [1]{%
 \ifx #1\expandafter \@firstoftwo
 \else \expandafter \@secondoftwo
 \fi
}%
\providecommand \natexlab [1]{#1}%
\providecommand \enquote  [1]{``#1''}%
\providecommand \bibnamefont  [1]{#1}%
\providecommand \bibfnamefont [1]{#1}%
\providecommand \citenamefont [1]{#1}%
\providecommand \href@noop [0]{\@secondoftwo}%
\providecommand \href [0]{\begingroup \@sanitize@url \@href}%
\providecommand \@href[1]{\@@startlink{#1}\@@href}%
\providecommand \@@href[1]{\endgroup#1\@@endlink}%
\providecommand \@sanitize@url [0]{\catcode `\\12\catcode `\$12\catcode
  `\&12\catcode `\#12\catcode `\^12\catcode `\_12\catcode `\%12\relax}%
\providecommand \@@startlink[1]{}%
\providecommand \@@endlink[0]{}%
\providecommand \url  [0]{\begingroup\@sanitize@url \@url }%
\providecommand \@url [1]{\endgroup\@href {#1}{\urlprefix }}%
\providecommand \urlprefix  [0]{URL }%
\providecommand \Eprint [0]{\href }%
\providecommand \doibase [0]{http://dx.doi.org/}%
\providecommand \selectlanguage [0]{\@gobble}%
\providecommand \bibinfo  [0]{\@secondoftwo}%
\providecommand \bibfield  [0]{\@secondoftwo}%
\providecommand \translation [1]{[#1]}%
\providecommand \BibitemOpen [0]{}%
\providecommand \bibitemStop [0]{}%
\providecommand \bibitemNoStop [0]{.\EOS\space}%
\providecommand \EOS [0]{\spacefactor3000\relax}%
\providecommand \BibitemShut  [1]{\csname bibitem#1\endcsname}%
\let\auto@bib@innerbib\@empty
\bibitem [{\citenamefont {Kippenberg}\ and\ \citenamefont
  {Vahala}(2008)}]{Kippenberg2008}%
  \BibitemOpen
  \bibfield  {author} {\bibinfo {author} {\bibfnamefont {T.~J.}\ \bibnamefont
  {Kippenberg}}\ and\ \bibinfo {author} {\bibfnamefont {K.~J.}\ \bibnamefont
  {Vahala}},\ }\href {http://www.sciencemag.org/content/321/5893/1172.full}
  {\bibfield  {journal} {\bibinfo  {journal} {Science}\ }\textbf {\bibinfo
  {volume} {321}},\ \bibinfo {pages} {1172} (\bibinfo {year}
  {2008})}\BibitemShut {NoStop}%
\bibitem [{\citenamefont {Marquardt}\ and\ \citenamefont
  {Girvin}(2009)}]{Marquardt2009}%
  \BibitemOpen
  \bibfield  {author} {\bibinfo {author} {\bibfnamefont {F.}~\bibnamefont
  {Marquardt}}\ and\ \bibinfo {author} {\bibfnamefont {S.~M.}\ \bibnamefont
  {Girvin}},\ }\href {http://physics.aps.org/articles/v2/40} {\bibfield
  {journal} {\bibinfo  {journal} {Physics}\ }\textbf {\bibinfo {volume} {2}},\
  \bibinfo {pages} {40} (\bibinfo {year} {2009})}\BibitemShut {NoStop}%
\bibitem [{\citenamefont {Milburn}\ and\ \citenamefont
  {Woolley}(2011)}]{Milburn2011}%
  \BibitemOpen
  \bibfield  {author} {\bibinfo {author} {\bibfnamefont {G.~J.}\ \bibnamefont
  {Milburn}}\ and\ \bibinfo {author} {\bibfnamefont {M.~J.}\ \bibnamefont
  {Woolley}},\ }\href {\doibase 10.2478/v10155-011-0005-7} {\bibfield
  {journal} {\bibinfo  {journal} {acta physica slovaca}\ }\textbf {\bibinfo
  {volume} {61}},\ \bibinfo {pages} {483} (\bibinfo {year} {2011})}\BibitemShut
  {NoStop}%
\bibitem [{\citenamefont {Aspelmeyer}\ \emph {et~al.}(2012)\citenamefont
  {Aspelmeyer}, \citenamefont {Meystre},\ and\ \citenamefont
  {Schwab}}]{Aspelmeyer2012}%
  \BibitemOpen
  \bibfield  {author} {\bibinfo {author} {\bibfnamefont {M.}~\bibnamefont
  {Aspelmeyer}}, \bibinfo {author} {\bibfnamefont {P.}~\bibnamefont {Meystre}},
  \ and\ \bibinfo {author} {\bibfnamefont {K.}~\bibnamefont {Schwab}},\ }\href
  {\doibase 10.1063/PT.3.1640} {\bibfield  {journal} {\bibinfo  {journal}
  {Physics Today}\ }\textbf {\bibinfo {volume} {65}},\ \bibinfo {pages} {29}
  (\bibinfo {year} {2012})}\BibitemShut {NoStop}%
\bibitem [{\citenamefont {Krishnan}\ \emph {et~al.}(2008)\citenamefont
  {Krishnan}, \citenamefont {Kshirsagar}, \citenamefont {Ananthasuresh},\ and\
  \citenamefont {Bhat}}]{Krishnan2007}%
  \BibitemOpen
  \bibfield  {author} {\bibinfo {author} {\bibfnamefont {G.}~\bibnamefont
  {Krishnan}}, \bibinfo {author} {\bibfnamefont {C.~U.}\ \bibnamefont
  {Kshirsagar}}, \bibinfo {author} {\bibfnamefont {G.~K.}\ \bibnamefont
  {Ananthasuresh}}, \ and\ \bibinfo {author} {\bibfnamefont {N.}~\bibnamefont
  {Bhat}},\ }\href@noop {} {\bibfield  {journal} {\bibinfo  {journal} {J.\
  Indian Inst.\ Sci.}\ }\textbf {\bibinfo {volume} {87}},\ \bibinfo {pages}
  {333} (\bibinfo {year} {2008})}\BibitemShut {NoStop}%
\bibitem [{\citenamefont {Binnig}\ \emph {et~al.}(1986)\citenamefont {Binnig},
  \citenamefont {Quate},\ and\ \citenamefont {Gerber}}]{Binning1986}%
  \BibitemOpen
  \bibfield  {author} {\bibinfo {author} {\bibfnamefont {G.}~\bibnamefont
  {Binnig}}, \bibinfo {author} {\bibfnamefont {C.~F.}\ \bibnamefont {Quate}}, \
  and\ \bibinfo {author} {\bibfnamefont {C.}~\bibnamefont {Gerber}},\ }\href
  {\doibase 10.1103/PhysRevLett.56.930} {\bibfield  {journal} {\bibinfo
  {journal} {Phys. Rev. Lett.}\ }\textbf {\bibinfo {volume} {56}},\ \bibinfo
  {pages} {930} (\bibinfo {year} {1986})}\BibitemShut {NoStop}%
\bibitem [{\citenamefont {Braginsky}\ and\ \citenamefont
  {Khalili}(1992)}]{Braginsky1992}%
  \BibitemOpen
  \bibfield  {author} {\bibinfo {author} {\bibfnamefont {V.~B.}\ \bibnamefont
  {Braginsky}}\ and\ \bibinfo {author} {\bibfnamefont {F.~Y.}\ \bibnamefont
  {Khalili}},\ }\href
  {http://www.cambridge.org/gb/knowledge/isbn/item1147478/?site_locale=en_GB}
  {\emph {\bibinfo {title} {Quantum Measurement}}}\ (\bibinfo  {publisher}
  {Cambridge University Press},\ \bibinfo {year} {1992})\BibitemShut {NoStop}%
\bibitem [{\citenamefont {Caves}\ \emph {et~al.}(1980)\citenamefont {Caves},
  \citenamefont {Thorne}, \citenamefont {Drever}, \citenamefont {Sandberg},\
  and\ \citenamefont {Zimmermann}}]{Caves1980}%
  \BibitemOpen
  \bibfield  {author} {\bibinfo {author} {\bibfnamefont {C.~M.}\ \bibnamefont
  {Caves}}, \bibinfo {author} {\bibfnamefont {K.~S.}\ \bibnamefont {Thorne}},
  \bibinfo {author} {\bibfnamefont {R.~W.~P.}\ \bibnamefont {Drever}}, \bibinfo
  {author} {\bibfnamefont {V.~D.}\ \bibnamefont {Sandberg}}, \ and\ \bibinfo
  {author} {\bibfnamefont {M.}~\bibnamefont {Zimmermann}},\ }\href {\doibase
  10.1103/RevModPhys.52.341} {\bibfield  {journal} {\bibinfo  {journal} {Rev.
  Mod. Phys.}\ }\textbf {\bibinfo {volume} {52}},\ \bibinfo {pages} {341}
  (\bibinfo {year} {1980})}\BibitemShut {NoStop}%
\bibitem [{\citenamefont {Clerk}\ \emph {et~al.}(2010)\citenamefont {Clerk}
  \emph {et~al.}}]{Clerk2010}%
  \BibitemOpen
  \bibfield  {author} {\bibinfo {author} {\bibfnamefont {A.~A.}\ \bibnamefont
  {Clerk}} \emph {et~al.},\ }\href
  {http://rmp.aps.org/abstract/RMP/v82/i2/p1155_1} {\bibfield  {journal}
  {\bibinfo  {journal} {Rev.\ Mod.\ Phys.}\ }\textbf {\bibinfo {volume} {82}},\
  \bibinfo {pages} {1155} (\bibinfo {year} {2010})}\BibitemShut {NoStop}%
\bibitem [{\citenamefont {Sheard}\ \emph {et~al.}(2004)\citenamefont {Sheard},
  \citenamefont {Gray}, \citenamefont {Mow-Lowry},\ and\ \citenamefont
  {McClelland}}]{Sheard2004}%
  \BibitemOpen
  \bibfield  {author} {\bibinfo {author} {\bibfnamefont {B.~S.}\ \bibnamefont
  {Sheard}}, \bibinfo {author} {\bibfnamefont {M.~B.}\ \bibnamefont {Gray}},
  \bibinfo {author} {\bibfnamefont {C.~M.}\ \bibnamefont {Mow-Lowry}}, \ and\
  \bibinfo {author} {\bibfnamefont {D.~E.}\ \bibnamefont {McClelland}},\ }\href
  {http://pra.aps.org/abstract/PRA/v69/i5/e051801} {\bibfield  {journal}
  {\bibinfo  {journal} {Phys.\ Rev.\ A}\ }\textbf {\bibinfo {volume} {69}},\
  \bibinfo {pages} {051801(R)} (\bibinfo {year} {2004})}\BibitemShut {NoStop}%
\bibitem [{\citenamefont {Arcizet}\ \emph
  {et~al.}(2006{\natexlab{a}})\citenamefont {Arcizet}, \citenamefont {Cohadon},
  \citenamefont {Briant}, \citenamefont {Pinard},\ and\ \citenamefont
  {Heidmann}}]{Arcizet2006nature}%
  \BibitemOpen
  \bibfield  {author} {\bibinfo {author} {\bibfnamefont {O.}~\bibnamefont
  {Arcizet}}, \bibinfo {author} {\bibfnamefont {P.-F.}\ \bibnamefont
  {Cohadon}}, \bibinfo {author} {\bibfnamefont {T.}~\bibnamefont {Briant}},
  \bibinfo {author} {\bibfnamefont {M.}~\bibnamefont {Pinard}}, \ and\ \bibinfo
  {author} {\bibfnamefont {A.}~\bibnamefont {Heidmann}},\ }\href
  {http://www.nature.com/nature/journal/v444/n7115/abs/nature05244.html}
  {\bibfield  {journal} {\bibinfo  {journal} {Nature}\ }\textbf {\bibinfo
  {volume} {444}},\ \bibinfo {pages} {71} (\bibinfo {year}
  {2006}{\natexlab{a}})}\BibitemShut {NoStop}%
\bibitem [{\citenamefont {Schliesser}\ \emph {et~al.}(2008)\citenamefont
  {Schliesser}, \citenamefont {Rivi\`{e}re}, \citenamefont {Anetsberger},
  \citenamefont {Arcizet},\ and\ \citenamefont {Kippenberg}}]{Schliesser2008}%
  \BibitemOpen
  \bibfield  {author} {\bibinfo {author} {\bibfnamefont {A.}~\bibnamefont
  {Schliesser}}, \bibinfo {author} {\bibfnamefont {R.}~\bibnamefont
  {Rivi\`{e}re}}, \bibinfo {author} {\bibfnamefont {G.}~\bibnamefont
  {Anetsberger}}, \bibinfo {author} {\bibfnamefont {O.}~\bibnamefont
  {Arcizet}}, \ and\ \bibinfo {author} {\bibfnamefont {T.~J.}\ \bibnamefont
  {Kippenberg}},\ }\href
  {http://www.nature.com/nphys/journal/v4/n5/abs/nphys939.html} {\bibfield
  {journal} {\bibinfo  {journal} {Nature Physics}\ }\textbf {\bibinfo {volume}
  {4}},\ \bibinfo {pages} {415} (\bibinfo {year} {2008})}\BibitemShut {NoStop}%
\bibitem [{\citenamefont {Bagheri}\ \emph {et~al.}(2011)\citenamefont
  {Bagheri}, \citenamefont {Poot}, \citenamefont {Li}, \citenamefont
  {Pernice},\ and\ \citenamefont {Tang}}]{Bagheri2011}%
  \BibitemOpen
  \bibfield  {author} {\bibinfo {author} {\bibfnamefont {M.}~\bibnamefont
  {Bagheri}}, \bibinfo {author} {\bibfnamefont {M.}~\bibnamefont {Poot}},
  \bibinfo {author} {\bibfnamefont {M.}~\bibnamefont {Li}}, \bibinfo {author}
  {\bibfnamefont {W.~P.~H.}\ \bibnamefont {Pernice}}, \ and\ \bibinfo {author}
  {\bibfnamefont {H.~X.}\ \bibnamefont {Tang}},\ }\href
  {http://www.nature.com/nnano/journal/v6/n11/full/nnano.2011.180.html}
  {\bibfield  {journal} {\bibinfo  {journal} {Nat.\ Nano.}\ }\textbf {\bibinfo
  {volume} {6}},\ \bibinfo {pages} {726} (\bibinfo {year} {2011})}\BibitemShut
  {NoStop}%
\bibitem [{\citenamefont {Xiao}\ \emph {et~al.}(1987)\citenamefont {Xiao},
  \citenamefont {Wu},\ and\ \citenamefont {Kimble}}]{Xiao1987}%
  \BibitemOpen
  \bibfield  {author} {\bibinfo {author} {\bibfnamefont {M.}~\bibnamefont
  {Xiao}}, \bibinfo {author} {\bibfnamefont {L.-A.}\ \bibnamefont {Wu}}, \ and\
  \bibinfo {author} {\bibfnamefont {H.~J.}\ \bibnamefont {Kimble}},\ }\href
  {\doibase 10.1103/PhysRevLett.59.278} {\bibfield  {journal} {\bibinfo
  {journal} {Phys. Rev. Lett.}\ }\textbf {\bibinfo {volume} {59}},\ \bibinfo
  {pages} {278} (\bibinfo {year} {1987})}\BibitemShut {NoStop}%
\bibitem [{\citenamefont {Thorne}\ \emph {et~al.}(1978)\citenamefont {Thorne},
  \citenamefont {Drever}, \citenamefont {Caves}, \citenamefont {Zimmermann},\
  and\ \citenamefont {Sandberg}}]{Thorne1978}%
  \BibitemOpen
  \bibfield  {author} {\bibinfo {author} {\bibfnamefont {K.~S.}\ \bibnamefont
  {Thorne}}, \bibinfo {author} {\bibfnamefont {R.~W.~P.}\ \bibnamefont
  {Drever}}, \bibinfo {author} {\bibfnamefont {C.~M.}\ \bibnamefont {Caves}},
  \bibinfo {author} {\bibfnamefont {M.}~\bibnamefont {Zimmermann}}, \ and\
  \bibinfo {author} {\bibfnamefont {V.~D.}\ \bibnamefont {Sandberg}},\ }\href
  {\doibase 10.1103/PhysRevLett.40.667} {\bibfield  {journal} {\bibinfo
  {journal} {Phys. Rev. Lett.}\ }\textbf {\bibinfo {volume} {40}},\ \bibinfo
  {pages} {667} (\bibinfo {year} {1978})}\BibitemShut {NoStop}%
\bibitem [{\citenamefont {Arcizet}\ \emph
  {et~al.}(2006{\natexlab{b}})\citenamefont {Arcizet} \emph
  {et~al.}}]{Arcizet2006}%
  \BibitemOpen
  \bibfield  {author} {\bibinfo {author} {\bibfnamefont {O.}~\bibnamefont
  {Arcizet}} \emph {et~al.},\ }\href
  {http://pra.aps.org/abstract/PRA/v73/i3/e033819} {\bibfield  {journal}
  {\bibinfo  {journal} {Phys.\ Rev.\ A}\ }\textbf {\bibinfo {volume} {73}},\
  \bibinfo {pages} {033819} (\bibinfo {year} {2006}{\natexlab{b}})}\BibitemShut
  {NoStop}%
\bibitem [{\citenamefont {Thompson}\ \emph {et~al.}(2008)\citenamefont
  {Thompson} \emph {et~al.}}]{Thompson2008}%
  \BibitemOpen
  \bibfield  {author} {\bibinfo {author} {\bibfnamefont {J.~D.}\ \bibnamefont
  {Thompson}} \emph {et~al.},\ }\href
  {http://www.nature.com/nature/journal/v452/n7183/abs/nature06715.html}
  {\bibfield  {journal} {\bibinfo  {journal} {Nature}\ }\textbf {\bibinfo
  {volume} {452}},\ \bibinfo {pages} {72} (\bibinfo {year} {2008})}\BibitemShut
  {NoStop}%
\bibitem [{\citenamefont {Jayich}\ \emph {et~al.}(2008)\citenamefont {Jayich}
  \emph {et~al.}}]{Jayich2008}%
  \BibitemOpen
  \bibfield  {author} {\bibinfo {author} {\bibfnamefont {A.~M.}\ \bibnamefont
  {Jayich}} \emph {et~al.},\ }\href
  {http://iopscience.iop.org/1367-2630/10/9/095008/} {\bibfield  {journal}
  {\bibinfo  {journal} {New \ J.\ Phys.}\ }\textbf {\bibinfo {volume} {10}},\
  \bibinfo {pages} {095008} (\bibinfo {year} {2008})}\BibitemShut {NoStop}%
\bibitem [{\citenamefont {Marquardt}(2011)}]{Marquardt2011}%
  \BibitemOpen
  \bibfield  {author} {\bibinfo {author} {\bibfnamefont {F.}~\bibnamefont
  {Marquardt}},\ }\href@noop {} {\enquote {\bibinfo {title} {Quantum
  optomechanics},}\ } (\bibinfo {year} {2011}),\ \bibinfo {note} {les Houches
  Lecture Notes 2011, School on ``Quantum Machines"}\BibitemShut {NoStop}%
\bibitem [{\citenamefont {Agarwal}(1984)}]{Agarwal1984}%
  \BibitemOpen
  \bibfield  {author} {\bibinfo {author} {\bibfnamefont {G.~S.}\ \bibnamefont
  {Agarwal}},\ }\href {http://prl.aps.org/abstract/PRL/v53/i18/p1732_1}
  {\bibfield  {journal} {\bibinfo  {journal} {Phys.\ Rev.\ Lett.}\ }\textbf
  {\bibinfo {volume} {53}},\ \bibinfo {pages} {1732} (\bibinfo {year}
  {1984})}\BibitemShut {NoStop}%
\bibitem [{\citenamefont {Thompson}\ \emph {et~al.}(1992)\citenamefont
  {Thompson}, \citenamefont {Rempe},\ and\ \citenamefont
  {Kimble}}]{Thompson1992}%
  \BibitemOpen
  \bibfield  {author} {\bibinfo {author} {\bibfnamefont {R.~J.}\ \bibnamefont
  {Thompson}}, \bibinfo {author} {\bibfnamefont {G.}~\bibnamefont {Rempe}}, \
  and\ \bibinfo {author} {\bibfnamefont {H.~J.}\ \bibnamefont {Kimble}},\
  }\href {http://prl.aps.org/abstract/PRL/v68/i8/p1132_1} {\bibfield  {journal}
  {\bibinfo  {journal} {Phys.\ Rev.\ Lett.}\ }\textbf {\bibinfo {volume}
  {68}},\ \bibinfo {pages} {1132} (\bibinfo {year} {1992})}\BibitemShut
  {NoStop}%
\bibitem [{\citenamefont {Bhattacharya}\ \emph {et~al.}(2008)\citenamefont
  {Bhattacharya}, \citenamefont {Uys},\ and\ \citenamefont
  {Meystre}}]{Bhattacharya2008}%
  \BibitemOpen
  \bibfield  {author} {\bibinfo {author} {\bibfnamefont {M.}~\bibnamefont
  {Bhattacharya}}, \bibinfo {author} {\bibfnamefont {H.}~\bibnamefont {Uys}}, \
  and\ \bibinfo {author} {\bibfnamefont {P.}~\bibnamefont {Meystre}},\ }\href
  {http://pra.aps.org/abstract/PRA/v77/i3/e033819} {\bibfield  {journal}
  {\bibinfo  {journal} {Phys.\ Rev.\ A}\ }\textbf {\bibinfo {volume} {77}},\
  \bibinfo {pages} {033819} (\bibinfo {year} {2008})}\BibitemShut {NoStop}%
\bibitem [{\citenamefont {Scully}\ and\ \citenamefont
  {Zubairy}(1997)}]{Scully1997}%
  \BibitemOpen
  \bibfield  {author} {\bibinfo {author} {\bibfnamefont {M.~O.}\ \bibnamefont
  {Scully}}\ and\ \bibinfo {author} {\bibfnamefont {M.~S.}\ \bibnamefont
  {Zubairy}},\ }\href
  {http://www.cambridge.org/gb/knowledge/isbn/item1143161/?site_locale=en_GB}
  {\emph {\bibinfo {title} {Quantum Optics}}}\ (\bibinfo  {publisher}
  {Cambridge University Press},\ \bibinfo {year} {1997})\BibitemShut {NoStop}%
\bibitem [{\citenamefont {Gardiner}\ and\ \citenamefont
  {Zoller}(2004)}]{Gardiner2004}%
  \BibitemOpen
  \bibfield  {author} {\bibinfo {author} {\bibfnamefont {C.~W.}\ \bibnamefont
  {Gardiner}}\ and\ \bibinfo {author} {\bibfnamefont {P.}~\bibnamefont
  {Zoller}},\ }\href
  {http://www.springer.com/physics/quantum+physics/book/978-3-540-22301-6}
  {\emph {\bibinfo {title} {Quantum Noise}}},\ \bibinfo {edition} {3rd}\ ed.\
  (\bibinfo  {publisher} {Springer, Berlin},\ \bibinfo {year}
  {2004})\BibitemShut {NoStop}%
\bibitem [{\citenamefont {Walls}\ and\ \citenamefont
  {Milburn}(2008)}]{Walls2008}%
  \BibitemOpen
  \bibfield  {author} {\bibinfo {author} {\bibfnamefont {D.}~\bibnamefont
  {Walls}}\ and\ \bibinfo {author} {\bibfnamefont {G.~J.}\ \bibnamefont
  {Milburn}},\ }\href
  {http://www.springer.com/physics/optics+\%26+lasers/book/978-3-540-28573-1}
  {\emph {\bibinfo {title} {Quantum Optics}}},\ \bibinfo {edition} {2nd}\ ed.\
  (\bibinfo  {publisher} {Springer},\ \bibinfo {year} {2008})\BibitemShut
  {NoStop}%
\bibitem [{\citenamefont {Kubo}(1966)}]{Kubo1966}%
  \BibitemOpen
  \bibfield  {author} {\bibinfo {author} {\bibfnamefont {R.}~\bibnamefont
  {Kubo}},\ }\href {http://iopscience.iop.org/0034-4885/29/1/306/} {\bibfield
  {journal} {\bibinfo  {journal} {Rep.\ Prog.\ Phys.}\ }\textbf {\bibinfo
  {volume} {29}},\ \bibinfo {pages} {255} (\bibinfo {year} {1966})}\BibitemShut
  {NoStop}%
\bibitem [{\citenamefont {DeJesus}\ and\ \citenamefont
  {Kaufman}(1987)}]{DeJesus1987}%
  \BibitemOpen
  \bibfield  {author} {\bibinfo {author} {\bibfnamefont {E.~X.}\ \bibnamefont
  {DeJesus}}\ and\ \bibinfo {author} {\bibfnamefont {C.}~\bibnamefont
  {Kaufman}},\ }\href {http://pra.aps.org/abstract/PRA/v35/i12/p5288_1}
  {\bibfield  {journal} {\bibinfo  {journal} {Phys.\ Rev.\ A}\ }\textbf
  {\bibinfo {volume} {35}},\ \bibinfo {pages} {5288} (\bibinfo {year}
  {1987})}\BibitemShut {NoStop}%
\bibitem [{\citenamefont {Genes}\ \emph {et~al.}(2008)\citenamefont {Genes}
  \emph {et~al.}}]{Genes2008}%
  \BibitemOpen
  \bibfield  {author} {\bibinfo {author} {\bibfnamefont {C.}~\bibnamefont
  {Genes}} \emph {et~al.},\ }\href
  {http://pra.aps.org/abstract/PRA/v77/i3/e033804} {\bibfield  {journal}
  {\bibinfo  {journal} {Phys.\ Rev.\ A}\ }\textbf {\bibinfo {volume} {77}},\
  \bibinfo {pages} {033804} (\bibinfo {year} {2008})}\BibitemShut {NoStop}%
\bibitem [{\citenamefont {Jeffrey}\ and\ \citenamefont
  {Zwillinger}(2000)}]{Routh-Hurwitz}%
  \BibitemOpen
  \bibfield  {author} {\bibinfo {author} {\bibfnamefont {A.}~\bibnamefont
  {Jeffrey}}\ and\ \bibinfo {author} {\bibfnamefont {D.}~\bibnamefont
  {Zwillinger}},\ }\href {http://dx.doi.org/10.1016/B978-012294757-5/50000-3}
  {\emph {\bibinfo {title} {Table of Integrals, Series and Products}}},\
  \bibinfo {edition} {sixth}\ ed.\ (\bibinfo  {publisher} {Academic Press},\
  \bibinfo {year} {2000})\BibitemShut {NoStop}%
\bibitem [{\citenamefont {Borkje}\ and\ \citenamefont
  {Girvin}(2012)}]{Borkje2012}%
  \BibitemOpen
  \bibfield  {author} {\bibinfo {author} {\bibfnamefont {K.}~\bibnamefont
  {Borkje}}\ and\ \bibinfo {author} {\bibfnamefont {S.~M.}\ \bibnamefont
  {Girvin}},\ }\href@noop {} {} (\bibinfo {year} {2012}),\ \Eprint
  {http://arxiv.org/abs/1204.6299v1} {arXiv:1204.6299v1} \BibitemShut {NoStop}%
\bibitem [{\citenamefont {Fermani}\ \emph {et~al.}(2004)\citenamefont
  {Fermani}, \citenamefont {Mancini},\ and\ \citenamefont
  {Tombesi}}]{Fermani2004}%
  \BibitemOpen
  \bibfield  {author} {\bibinfo {author} {\bibfnamefont {R.}~\bibnamefont
  {Fermani}}, \bibinfo {author} {\bibfnamefont {S.}~\bibnamefont {Mancini}}, \
  and\ \bibinfo {author} {\bibfnamefont {P.}~\bibnamefont {Tombesi}},\ }\href
  {\doibase 10.1103/PhysRevA.70.045801} {\bibfield  {journal} {\bibinfo
  {journal} {Phys.\ Rev.\ A}\ }\textbf {\bibinfo {volume} {70}},\ \bibinfo
  {pages} {045801} (\bibinfo {year} {2004})}\BibitemShut {NoStop}%
\bibitem [{\citenamefont {Kemiktarak}\ \emph
  {et~al.}(2012{\natexlab{a}})\citenamefont {Kemiktarak}, \citenamefont
  {Durand}, \citenamefont {Metcalfe},\ and\ \citenamefont
  {Lawall}}]{Kemiktarak2012}%
  \BibitemOpen
  \bibfield  {author} {\bibinfo {author} {\bibfnamefont {U.}~\bibnamefont
  {Kemiktarak}}, \bibinfo {author} {\bibfnamefont {M.}~\bibnamefont {Durand}},
  \bibinfo {author} {\bibfnamefont {M.}~\bibnamefont {Metcalfe}}, \ and\
  \bibinfo {author} {\bibfnamefont {J.}~\bibnamefont {Lawall}},\ }\href
  {http://stacks.iop.org/1367-2630/14/i=12/a=125010} {\bibfield  {journal}
  {\bibinfo  {journal} {New Journal of Physics}\ }\textbf {\bibinfo {volume}
  {14}},\ \bibinfo {pages} {125010} (\bibinfo {year}
  {2012}{\natexlab{a}})}\BibitemShut {NoStop}%
\bibitem [{\citenamefont {Collett}\ and\ \citenamefont
  {Gardiner}(1984)}]{Collett1984}%
  \BibitemOpen
  \bibfield  {author} {\bibinfo {author} {\bibfnamefont {M.~J.}\ \bibnamefont
  {Collett}}\ and\ \bibinfo {author} {\bibfnamefont {C.~W.}\ \bibnamefont
  {Gardiner}},\ }\href {\doibase 10.1103/PhysRevA.30.1386} {\bibfield
  {journal} {\bibinfo  {journal} {Phys.\ Rev.\ A}\ }\textbf {\bibinfo {volume}
  {30}},\ \bibinfo {pages} {1386} (\bibinfo {year} {1984})}\BibitemShut
  {NoStop}%
\bibitem [{\citenamefont {Stannigel}\ \emph {et~al.}(2011)\citenamefont
  {Stannigel}, \citenamefont {Rabl}, \citenamefont {S\o{}rensen}, \citenamefont
  {Lukin},\ and\ \citenamefont {Zoller}}]{PeterRabl2011}%
  \BibitemOpen
  \bibfield  {author} {\bibinfo {author} {\bibfnamefont {K.}~\bibnamefont
  {Stannigel}}, \bibinfo {author} {\bibfnamefont {P.}~\bibnamefont {Rabl}},
  \bibinfo {author} {\bibfnamefont {A.~S.}\ \bibnamefont {S\o{}rensen}},
  \bibinfo {author} {\bibfnamefont {M.~D.}\ \bibnamefont {Lukin}}, \ and\
  \bibinfo {author} {\bibfnamefont {P.}~\bibnamefont {Zoller}},\ }\href
  {\doibase 10.1103/PhysRevA.84.042341} {\bibfield  {journal} {\bibinfo
  {journal} {Phys.\ Rev.\ A}\ }\textbf {\bibinfo {volume} {84}},\ \bibinfo
  {pages} {042341} (\bibinfo {year} {2011})}\BibitemShut {NoStop}%
\bibitem [{\citenamefont {Kemiktarak}\ \emph
  {et~al.}(2012{\natexlab{b}})\citenamefont {Kemiktarak}, \citenamefont
  {Metcalfe}, \citenamefont {Durand},\ and\ \citenamefont
  {Lawall}}]{Kemiktarak2012apl}%
  \BibitemOpen
  \bibfield  {author} {\bibinfo {author} {\bibfnamefont {U.}~\bibnamefont
  {Kemiktarak}}, \bibinfo {author} {\bibfnamefont {M.}~\bibnamefont
  {Metcalfe}}, \bibinfo {author} {\bibfnamefont {M.}~\bibnamefont {Durand}}, \
  and\ \bibinfo {author} {\bibfnamefont {J.}~\bibnamefont {Lawall}},\ }\href
  {http://apl.aip.org/resource/1/applab/v100/i6/p061124_s1} {\bibfield
  {journal} {\bibinfo  {journal} {Appl.\ Phys.\ Lett.}\ }\textbf {\bibinfo
  {volume} {100}},\ \bibinfo {pages} {061124} (\bibinfo {year}
  {2012}{\natexlab{b}})}\BibitemShut {NoStop}%
\end{thebibliography}%

\newpage
\onecolumngrid
\section{\large Supplemental Information}
\section{Normal mode splitting in the high reflectivity limit}
The hamiltonian of the cavity fields in this three-mirror system is
\allowdisplaybreaks
\begin{eqnarray}
H_{\mathrm{opt}}  & = &   \hbar (\omega_c +   f  x )  a_L^{\dagger} a_L  + \hbar  (\omega_c  -  f   x ) a_R^{\dagger} a_R  - \hbar g(a_L^{\dagger}a_R + a_R^{\dagger}a_L),
\end{eqnarray}
Diagonalizing this coupled left/right mode hamiltonian, we find two new eigenfrequencies:  
\begin{equation}
\omega_{cav, \pm} (x) = \omega_c  \pm \sqrt{f^2  x^2 + g^2},
\label{splitting}
\end{equation} 
which correspond to the normal mode frequencies of the three-mirror system, and the difference between these two eigenfrequencies is exactly the normal mode splitting in Eq.~(\ref{splitting}),
\begin{equation}
2\sqrt{f^2 x^2 + g^2} =  \frac{c}{L}\arccos(|r_d|\cos(2kx)). 
\end{equation} 
Matching the left and right hand sides at a small displacement range, we find $g = \arccos(|r_d|) c/2L$ and $f = - \sqrt{|r_d| \arcsin(|t_d|)/|t_d|}\omega_c/L$. In the high reflectivity limit ($|r_d| \rightarrow 1$, $|t_d| \rightarrow 0$), $g = |t_d| c/2L$ and $ f = - \omega_c/L$, which implies that the coupling constant is proportional to the transmission of the middle mirror and that the dispersion of the cavity resonant frequency is linear. We have to mention that the full expression for the resonant frequency shift $f$ is valid at both high and moderate reflectivity, as soon as the displacement $x$ is small compared with the wavelength of the cavity field. But problem rises at moderate reflectivity, because the normal mode splitting $\Omega$ increases as reflectivity decreases, and it approaches the free spectrum range of the large cavity with length $2L$. To avoid the coupling between multiple cavity modes, we only consider the high reflectivity case, which has been realized experimentally \cite{Kemiktarak2012apl}.

To simplify the hamiltonian, we define two new modes  $a= (a_L + a_R)/\sqrt{2}$ and $b= (a_L - a_R)/\sqrt{2}$. If we drive mode $a$ strongly at frequency $\omega_L$, we can move to the rotating frame with respect to the pump laser, then we have the hamiltonian of the whole system (vibrating mirror, cavity modes and pump):
\begin{eqnarray}
H & = &   \frac{p^2}{2m} + \frac{1}{2}m\omega_m^2 x^2  + \hbar (\Delta_c -g) a^{\dagger} a +   \hbar (\Delta_c + g) b^{\dagger} b  - \hbar\omega_c \frac{x}{L} (a^{\dagger} b  +  b^{\dagger} a)  +  i\hbar E(a ^{\dagger}  - a ),
\end{eqnarray}
where $\Delta_c = \omega_c - \omega_L$ is the cavity detuning, and $E$ is the pump strength which is related to input laser power $P_{in}$  and cavity damping $\kappa$  by $|E| = \sqrt{P_{in}\kappa/\hbar\omega_L}$.

\section{Solve the equations of motion}
To solve the reduced Heisenberg-Langevin equations, we define the quadratures of mode $b$ as $X =  (b + b^{\dagger})/\sqrt{2}$,  $Y = (b - b^{\dagger})/i \sqrt{2}$, and move to frequency domain by Fourier transform, obtaining the following equations of motion:
\begin{subequations}
\begin{eqnarray}
-i\omega x(\omega) & = & p(\omega)/m,  \\
-i\omega p(\omega) & = & -m\omega_m^2 x(\omega) - \gamma p(\omega) +  \frac{\sqrt{2} \hbar \omega_c \alpha}{L} X(\omega)
  + F_{in}(\omega),  \\
-i\omega X(\omega) & = & -\kappa X(\omega) + \Delta Y(\omega) +  \sqrt{2\kappa}X_{in}(\omega), \\
-i\omega Y(\omega) & = & -\kappa Y(\omega) - \Delta X(\omega) +  \frac{\sqrt{2}\omega_c \alpha}{L} x(\omega) + \sqrt{2\kappa} Y_{in}(\omega). 
\end{eqnarray}
\end{subequations}
From this set of coupled linear equations, we find, 
\begin{subequations}
\begin{eqnarray}
X(\omega) & = & \frac{1}{(\kappa-i\omega)^2 + \Delta^2} \left[\frac{\sqrt{2}\omega_c \alpha \Delta }{L} x(\omega)  + \sqrt{2\kappa} (\kappa-i\omega) X_{in}(\omega) + \sqrt{2\kappa} \Delta Y_{in}(\omega) \vphantom{\frac{\sqrt{2}\omega_c \alpha }{L}}\right],  \\
Y(\omega) & = & \frac{1}{(\kappa-i\omega)^2 + \Delta^2} \left[\frac{ \sqrt{2}\omega_c \alpha }{L} (\kappa-i\omega) x(\omega)  + \sqrt{2\kappa} (\kappa-i\omega) Y_{in}(\omega) - \sqrt{2\kappa} \Delta X_{in}(\omega) \vphantom{\frac{\sqrt{2}\omega_c \alpha }{L}} \right].
\end{eqnarray} 
\end{subequations}
At zero detuning $\Delta = 0$, $X(\omega)$ is unchanged, while $Y(\omega)$ is modulated by oscillator displacement $x(\omega)$ \cite{Kippenberg2008}. Measuring $Y(\omega)$ will give us the information about oscillator displacement.  At finite detuning, however, both quadratures are related to oscillator displacement.  Putting the field quadratures back into the equation of motion for the oscillator, we have
\begin{eqnarray}
x(\omega) &=& \chi(\omega) \left\{ F_{in}(\omega)  +  \cfrac{2\hbar G\sqrt{\kappa}}{(\kappa-i\omega)^2 +\Delta^2} \left [(\kappa-i\omega) X_{in}(\omega) + \Delta Y_{in}(\omega) \right]\right\},
\end{eqnarray}
with $G = \omega_c\alpha/L$ and $\chi(\omega)$ is the  susceptibility of the optomechanical system to force, 
\begin{equation}
\chi(\omega) = \left\{m\left [\omega_m^2 -  \omega^2 -i\gamma \omega  -\cfrac{2\hbar G^2\Delta/m}{(\kappa-i\omega)^2 +\Delta^2} \right] \right\} ^{-1} . 
\end{equation}

\section{Force detection sensitivity and optimization}
The vacuum radiation input noise $b_{in}$ is delta correlated and the thermal fluctuating force is approximated as white noise thus is also delta correlated, so we have the following correlation relations in frequency domain:\begin{subequations}
\begin{eqnarray}
\langle X_{in}(\omega) X_{in} (\omega^{\prime}) \rangle  & = &  \langle Y_{in}(\omega) Y_{in} (\omega^{\prime}) \rangle =  \frac{1}{2} \delta(\omega+\omega^{\prime}) , \\
\langle X_{in}(\omega) Y_{in} (\omega^{\prime}) \rangle & = &  -\langle Y_{in}(\omega) X_{in} (\omega^{\prime}) \rangle =   \frac{i}{2}  \delta(\omega+\omega^{\prime}) ,\\
\langle F_{th}(\omega)F_{th}(\omega^{\prime}) \rangle & = & 2m\gamma k_{B}T \delta(\omega + \omega^{\prime}), 
\end{eqnarray}
\label{correlation_relation}
\end{subequations} 
from which we have
\begin{eqnarray}
\eta(\omega) & =& 2m\gamma k_B T + \frac{1}{2} \left|\cfrac{\chi_X(\omega) - i\chi_Y(\omega)}{\chi_F(\omega)}\right|^2  \nonumber \\
&=&2m\gamma k_B T  +  \frac{1}{2} \left|\cfrac{2 \hbar\sqrt{\kappa}G}{\kappa-i\omega + i\Delta} + \cfrac{(\kappa - i\Delta)^2 + \omega^2}{2\sqrt{\kappa}G} \frac{\sin\theta - i\cos\theta}{\Delta\sin\theta + (\kappa - i\omega)\cos\theta} \chi^{-1}(\omega) \right|^2 \nonumber \\
&=& 2m\gamma k_B T + \cfrac{2 \hbar^2\kappa G^2}{|\kappa-i\omega + i\Delta|^2} + \cfrac{|(\kappa - i\Delta)^2 + \omega^2|^2}{8\kappa G^2} \frac{m^2|{\omega_m^{\prime}}^2 - \omega^2 - i\gamma^{\prime}\omega|^2}{|\Delta\sin\theta + (\kappa - i\omega)\cos\theta|^2}  \nonumber \\
&& + \frac{\hbar}{2} (\kappa + i\Delta + i\omega)  \frac{\sin\theta + i\cos\theta}{\Delta\sin\theta + (\kappa + i\omega)\cos\theta}m({\omega_m^{\prime}}^2 - \omega^2 + i\gamma^{\prime}\omega) + c.c.
\end{eqnarray} 
The first term is the thermal white noise. The second term proportional to $G^2$ comes the random back action force. The third term proportional to $1/G^2$ is the phase noise related to position measurement imprecision. 

In the vicinity of $\theta_0 = -\arctan(\kappa/\Delta)$, $\theta = \theta_0 + \delta\theta$, and the optimal pump is approximated as 
\begin{equation}
\alpha_{\ast}^2  = \alpha_0^2 (1 - \frac{2\kappa}{\Delta}\delta\theta),
\end{equation}
which leads to an effective mechanical frequency
\begin{equation}
{\omega_m^{\prime}}^2 = \omega_m^2 - \frac{2\hbar\Delta\omega_c^2\alpha^2/mL^2}{\kappa^2 + \Delta^2} = \omega_m^2 \frac{2\kappa}{\Delta}\delta\theta.
\end{equation}
The susceptibilities at DC reduce to 
\begin{subequations}
\begin{eqnarray}
|\chi_F|^2 &=& \frac{1}{2\hbar m\omega_m^2}\frac{\Delta}{\kappa} (1 - \frac{2\kappa}{\Delta}\delta\theta), \\
|\chi_X - i\chi_Y|^2 &=&  \left| (\frac{2\kappa}{\Delta} + i) \delta\theta \right|^2.
\end{eqnarray}
\end{subequations}
In the limit $\delta\theta \to 0$, the force susceptibility remains finite, while the optical susceptibilities go to zero. 
So the total sensitivity is given by 
\begin{equation}
\eta(\omega = 0) 
\approx   2m\gamma k_B T + \hbar m\omega_m^2 \left(\frac{\Delta}{4\kappa} + \frac{\kappa}{\Delta}\right)\xi^2.
\end{equation}

\end{document}